\documentclass[%
aps, twocolumn,
superscriptaddress,
bibnotes,
 amsmath,amssymb,
 aps,
numerical
]{revtex4-2}

\usepackage{natbib}

\usepackage{etoolbox}
\usepackage[utf8]{inputenc}
\usepackage{siunitx}
\usepackage{array}
\usepackage{physics}
\usepackage{dsfont}
\usepackage[english]{babel}			
\usepackage{xcolor}
\usepackage{graphicx}				
\usepackage{amssymb,amsmath}		
\usepackage{url}					
\usepackage{marvosym}				
\usepackage[T1]{fontenc}            %
\usepackage{wrapfig}				
\usepackage{charter} 				
\usepackage{blindtext}				
\usepackage{datetime}				
\usepackage{lipsum}                 
\usepackage{float}                  
\usepackage{tabularx}
\usepackage{dcolumn}
\usepackage{bm}
\usepackage{todonotes}
\usepackage{hyperref}
\hypersetup{colorlinks=true,allcolors=blue}

\newcommand{\UIBK}{Institut f{\"u}r Experimentalphysik, Universit{\"a}t Innsbruck, Innsbruck, Austria}
\newcommand{\WWU}{Institut f{\"u}r Festk{\"o}rpertheorie, Universit{\"a}t M{\"u}nster, 48149 M{\"u}nster, Germany}
\newcommand{\TUdo}{Condensed Matter Theory, Department of Physics, TU Dortmund, 44221 Dortmund, Germany}
\newcommand{\Bayreuth}{Theoretische Physik III, Universit{\"a}t Bayreuth, 95440 Bayreuth, Germany}
\newcommand{\JKU}{Institute of Semiconductor and Solid State Physics, Johannes Kepler University Linz, Linz, Austria}
\newcommand{\UniWien}{University of Vienna, Faculty of Physics, Vienna Center for Quantum Science and Technology (VCQ), Vienna, Austria, \\ Christian Doppler Laboratory for Photonic Quantum Computer, Faculty of Physics, University of Vienna, Vienna, Austria}
\newcommand{\TuB}{Institute of Solid State Physics, Technische Universität Berlin, 10623 Berlin, Germany}

\preprint{APS/123-QED}

\begin{document}

\title{Controlling the Photon Number Coherence of Solid-state Quantum Light Sources for Quantum Cryptography}

\author{Yusuf Karli}
\thanks{These authors contributed equally}
\affiliation{\UIBK}
\author{Daniel A. Vajner}
    \thanks{These authors contributed equally}
\affiliation{\TuB}
\author{Florian Kappe}
    \thanks{These authors contributed equally}
\affiliation{\UIBK}
\author{Paul C. A. Hagen}
\affiliation{\Bayreuth}
\author{Lena M. Hansen}
\affiliation{\UniWien}
\author{Ren\'e Schwarz}
\affiliation{\UIBK}
\author{Thomas K. Bracht}
\affiliation{\TUdo}
\affiliation{\WWU}
\author{Christian Schimpf}
\affiliation{\JKU}
\author{Saimon F. Covre da Silva}
\affiliation{\JKU}
\author{Philip Walther}
\affiliation{\UniWien}
\author{Armando Rastelli}
\affiliation{\JKU}
\author{Vollrath Martin Axt}
\affiliation{\Bayreuth}
\author{Juan C. Loredo}
\affiliation{\UniWien}
\author{Vikas Remesh}
\affiliation{\UIBK}
\author{Tobias Heindel}
\affiliation{\TuB}
\author{Doris E. Reiter}
\affiliation{\TUdo}
\author{Gregor Weihs}
\affiliation{\UIBK}

\makeatletter
\patchcmd{\frontmatter@RRAP@format}{(}{}{}{}
\patchcmd{\frontmatter@RRAP@format}{)}{}{}{}
\renewcommand\Dated@name{}
\makeatother

\date{Date: \today \\ \phantom{XXX} E-mail: yusuf.karli@uibk.ac.at}


\begin{abstract}
Quantum communication networks rely on quantum cryptographic protocols including quantum key distribution (QKD) using single photons. A critical element regarding the security of QKD protocols is the photon number coherence (PNC), i.e. the phase relation between the zero and one-photon Fock state, which critically depends on the excitation scheme. Thus, to obtain flying qubits with the desired properties, optimal pumping schemes for quantum emitters need to be selected. Semiconductor quantum dots generate on-demand single photons with high purity and indistinguishability. Exploiting two-photon excitation of a quantum dot combined with a stimulation pulse, we demonstrate the generation of high-quality single photons with a controllable degree of PNC. Our approach provides a viable route toward secure communication in quantum networks.
\end{abstract}

\maketitle

\section{Introduction}
Single photons are an essential resource for future high-security communication networks, with applications like measurement-based or distributed quantum computing and quantum cryptography \cite{gisin2007quantum,briegel2009measurement,flamini2018photonic}. Every quantum information protocol has its unique set of practical requirements \cite{pirandola2020advances}. While early quantum key distribution (QKD) protocols \cite{bennett1984proceedings,ekert1991quantum} primarily relied on high single-photon purity, more advanced schemes have further requirements such as high indistinguishability, for example in quantum repeaters or measurement-device-independent (MDI)-QKD, which relies on remote two-photon interference \cite{briegel1998quantum,lo2012measurement}. 
The search for efficient single-photon sources has led to semiconductor quantum dots \cite{vajner_quantum_2022}, thanks to their high single-photon purity \cite{hanschke_quantum_2018}, brightness \cite{tomm_bright_2021}, indistinguishability \cite{zhai2022quantum}, scalability \cite{uppu2020scalable} and above all, versatility in emission wavelength selection.

\begin{figure}[ht]
    \centering
    \includegraphics[width=\linewidth]{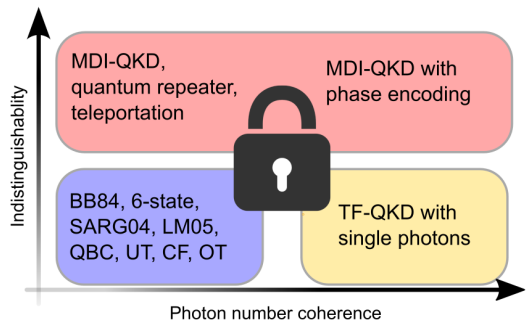}
    \caption{
     \textbf{Overview of various quantum information protocols:} The protocols are sorted by their requirements on indistinguishability $\cal{I}$ and PNC, specifically focusing on discrete variable cryptographic protocols using polarization or time-bin encoding. Protocols that require low PNC and need no high $\cal{I}$ are BB84 \cite{bennett1984proceedings}, decoy-BB84 \cite{wang2005beating}, 6-state-protocol \cite{bruss1998optimal}, SARG04 \cite{scarani2004quantum}, LM05 \cite{lucamarini2005secure}, QBC: quantum bit commitment \cite{ng2012experimental}, UT: unforgeable quantum tokens \cite{pastawski2012unforgeable,bozzio2018experimental}, CF: quantum coin flipping \cite{berlin2011experimental,pappa2014experimental}, OT: Oblivious Transfer \cite{bennett2001practical}; low PNC and high $\cal{I}$ is required by MDI-QKD \cite{lo2012measurement}, quantum repeaters and entanglement swapping for QKD \cite{briegel1998quantum,pan1998experimental}, quantum teleportation for QKD \cite{bouwmeester1997experimental}, DI-QKD with single photons \cite{gonzalez2022violation}; high PNC and variable $\cal{I}$ are needed in TF: twin-field QKD with single photons \cite{lucamarini2018overcoming}. Note that carrying out protocols from the low PNC column with phase encoding would require an initially defined phase, before randomizing it in a reversible way, which requires PNC in the beginning. To illustrate this we have also added MDI-QKD with phase encoding to the diagram.} 
    \label{fig:QKD}
\end{figure}

Photon number coherence (PNC) \cite{loredo2019generation, Wein2022} is another crucial quantity relevant for the security of single photon quantum cryptography schemes. It must vanish for most protocols \cite{bozzio2022enhancing}, compromising security otherwise \cite{gottesman2004security,lim2014concise} due to side-channel attacks enabled by the fixed relative phase between different photon number states \cite{lo2005phase,tang2013source}. While there are more general security proofs that allow a non-zero PNC, they lead to lower key rates, as some of the bits must be devoted to compensate for the additional information leakage towards an eavesdropper \cite{lo2006security}. To achieve zero PNC in practice, actively phase-randomized single photons can be used, as typically implemented for faint laser pulses \cite{zhao2007experimental,kobayashi2014evaluation}. Otherwise, a suitable excitation scheme without PNC must be chosen, which might however deteriorate the other single photon properties \cite{loredo2019generation,bozzio2022enhancing}.

In this work, we achieve tailored degrees of PNC, on demand, maintaining high purity and indistinguishability. We implement optical excitation protocols to demonstrate the experimental single-photon generation from quantum dots. The photon output in our scheme can be increased up to twice as high compared to more commonly used methods like resonant excitation. We therefore set the stage for the quantum dot platform to be used for advanced cryptographic implementations.

Due to its versatility, the excitation scheme presented here covers the requirements of a broad range of quantum cryptographic protocols. An overview of various applications in the context of PNC and indistinguishability requirements is given in Fig. \ref{fig:QKD}\textbf{a}. For example, established protocols like BB84 \cite{bennett1984proceedings}, decoy-BB84 \cite{wang2005beating}, 6-state-protocol \cite{bruss1998optimal}, SARG04 \cite{scarani2004quantum}, LM05 \cite{lucamarini2005secure} and primitives like strong quantum coin flipping \cite{berlin2011experimental,pappa2014experimental}, unforgeable quantum tokens \cite{pastawski2012unforgeable,bozzio2018experimental}, quantum bit commitment \cite{ng2012experimental} or quantum oblivious transfer \cite{santos2022quantum} require the absence of PNC to ensure, for instance, security in QKD or fairness in coin-flipping protocols. On the other hand, there exist protocols that benefit from a finite amount of initial PNC like MDI-QKD when done with phase encoding \cite{lo2012measurement} or twin-field QKD protocols \cite{lucamarini2018overcoming} 
to know and set the initial phase \cite{bozzio2022enhancing}.

\begin{figure*}[ht]
    \centering
    \includegraphics[width=1\linewidth]{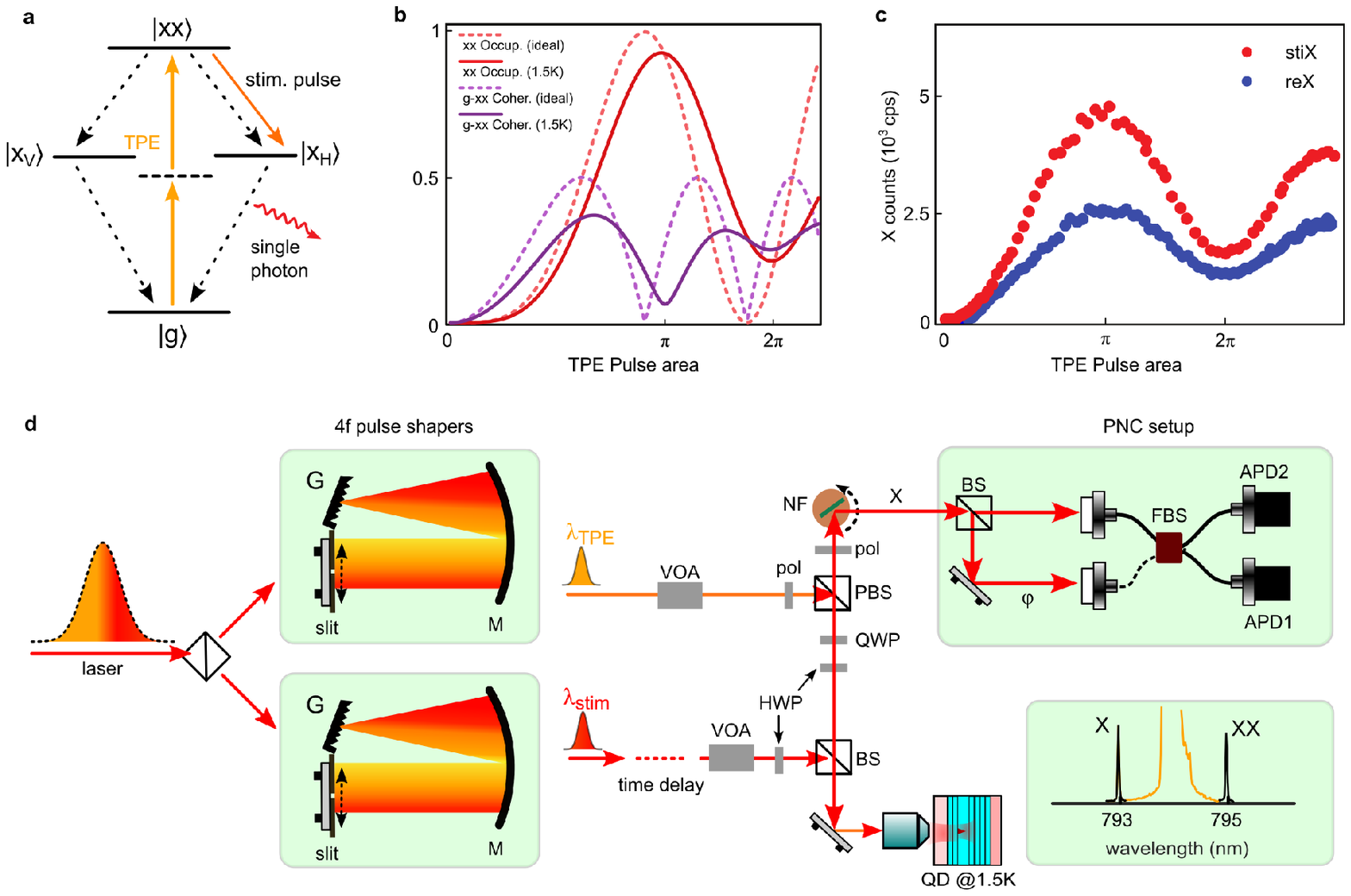}
    \caption{
    \textbf{Generating single photons with variable PNC:}
    (\textbf{a}) Level scheme of a quantum dot consisting of ground state $|g\rangle$, two linearly polarized exciton states $|x_{V/H}\rangle$ and biexciton state $|xx\rangle$. Straight lines indicate laser excitation, while dashed lines denote relaxation processes with rate $\gamma$. Both schemes start with a two-photon excitation (TPE) from $|g\rangle \to |xx\rangle$. In stiX, an additional H-polarized laser pulse stimulates the transition $|xx\rangle \to |x_H\rangle$. We only collect $H$-polarized photons. (\textbf{b}) Theoretically calculated biexciton ($|xx\rangle$) occupation showing Rabi rotations as a function of the TPE pulse area (red curve) and the corresponding coherence between $|g\rangle$ and $|xx\rangle$ (purple curve).  (\textbf{c}) Exciton photon counts recorded under reX (blue dots) and stiX (red dots) manifesting the enhancement of photon counts under stimulation. (\textbf{d}) Sketch of the experimental setup: a Ti:Sapphire laser source producing $\approx$\SI{2}{\pico\second}-long laser pulses, with a spectral FWHM of \SI{0.5}{\nano\meter}, is used to spectrally shape TPE and stim. pulses at appropriate wavelengths $\lambda_{\text{TPE}}$ and $\lambda_{\text{stim}}$ using two 4$f$ pulse shapers. A fiber-optic delay line enables the time control of the stim. pulse with respect to the TPE pulse. An electronic variable optical attenuator (VOA) helps sweep the laser power. The two pulses meet at a 10:90 beamsplitter (BS) and propagate to the cryostat which holds the quantum dot at \SI{1.5}{\kelvin}. Emitted single photons from the quantum dot are spectrally filtered by a notch filter (NF) and send to an unbalanced Mach-Zehnder interferometer with a freely evolving phase on one arm (labeled as PNC setup). Two single-photon sensitive avalanche photodiodes (APD1 and APD2) detect the single photon counts at the output arms of the interferometer. pol: linear polarizer, HWP: half-wave plate, QWP: quarter-wave plate, BS: beamsplitter, PBS: polarizing beam splitter, FBS: fiber beam splitter.}
    \label{fig:intro}
\end{figure*}

Our excitation protocols are based on resonant two-photon excitation (TPE) of a quantum dot from the ground state $|g\rangle$ into the biexciton state $|xx\rangle$ \cite{stufler_two-photon_2006,jayakumar_deterministic_2013}, yielding Rabi rotations. A simulation of TPE Rabi rotations is shown in Fig.~\ref{fig:intro}\textbf{b} (red dashed line) and experimental data in Fig.~\ref{fig:intro}\textbf{c} (blue dots). 
From the biexciton state, the system relaxes into either horizontally $|x_{H}\rangle$ or vertically $|x_{V}\rangle$ polarized exciton state, from which we collect only horizontally ($H$) polarized photons. 
We call this scheme \emph{relaxation into the exciton} (reX). 
The reX scheme is advantageous over the direct, resonant excitation of the exciton, due to the suppressed re-excitation and therefore provides high-purity photon states \cite{hanschke_quantum_2018}. 
Because the exciting laser energy is different from the emitted photon energy, a challenging cross-polarization filtering is avoided and the photon count rate can be increased up to a factor of two, which is also achieved by several other recently-proposed excitation schemes \cite{reindl2019highly,cosacchi2019emission,thomas_bright_2021,bracht2021swing,karli2022super,koong2021coherent,wilbur2022notch}. 
However, the indistinguishability of the single photons via the reX scheme suffers greatly from the spontaneous decay of the biexciton \cite{scholl2020crux} and if a specific polarization is required, the photon output is reduced due to the two available decay channels.

An improved protocol to overcome these problems uses an additional stimulation laser pulse following the TPE pulse \cite{akimov2006stimulated}. This \emph{stimulated preparation of the exciton} (stiX) scheme can generate higher indistinguishability exciton photons due to the reduced time jitter \cite{sbresny2022stimulated,yan2022double,wei2022tailoring}. Because the stimulation pulse determines the polarization of the emitted photon, the photon counts in that polarization state is also enhanced up to a factor of two (see also Fig.~\ref{fig:intro}\textbf{c}). Although the presence of PNC under resonant excitation and reX has been investigated before \cite{loredo2019generation,bozzio2022enhancing}, it remains to be seen if PNC exists in the stiX scheme. Additionally, assessing the controllability of PNC is essential for advancing optical preparation schemes of quantum dot states for quantum cryptography applications.

\section{Results}
\subsection*{Definition of photon number coherence}
In a pure state  $|\Psi\rangle =  \sum_{n=0}^{\infty} c_n|n\rangle$ in the photon number Fock basis with eigenstates $|n \rangle$ and the complex coefficients $c_n$, we define PNC as the absolute value of the coherence between the Fock states. For QKD based on single photons, as considered in this paper, the PNC refers to the coherence between the Fock states $|0\rangle$ and $|1\rangle$. More generally, we employ a density matrix description using
\begin{equation}
    \rho = \left(\begin{array}{cc}
         \rho_{0,0} & \rho_{0,1}  \\
         \rho_{1,0}& \rho_{1,1}
    \end{array} \right) \qquad \text{with} \qquad \text{PNC}=|\rho_{0,1}|,
\end{equation}
 $\rho_{1,1}$ ($\rho_{0,0}$) being the occupation of the one (zero)-photon state and $\rho_{0,1}$ being the coherence. We recall that it holds that $|\rho_{0,1}|^2 \le \rho_{1,1}\, \rho_{0,0}$ with equality in case of a pure state. The inequality implies that for $\rho_{1,1}=1$ or $\rho_{0,0}=1$ the PNC vanishes, while for $\rho_{0,0}=\rho_{1,1}=1/2$ it is maximal.

There are several factors that affect the PNC. One aspect is the non-perfect preparation of the photon state. The properties of photons from a quantum dot depend on the preparation fidelity of the quantum dot electronic state. This fidelity, in turn, is affected by the interaction with the environment of the quantum dot, most strongly by the interaction with phonons \cite{luker2019review}. Phonons can also degrade the coherence properties of the photons \cite{cosacchi2021accuracy} and therefore also the PNC. Additional losses of photons into other modes, which are not detected, further affect the photon properties. 

It is likewise important to consider the measurement process. To detect PNC, a phase-evolving Mach-Zehnder Interferometer (MZI) is employed \cite{loredo2019generation}. The outputs of the MZI are simultaneously recorded with two avalanche photodiodes (APDs). The count rates $N_1, N_2$ in the APD result in the visibility
\begin{equation}
    v_i = \frac{N_i^\text{max}-N_i^\text{min}}{N_i^\text{max}+N_i^\text{min}} \,.
\end{equation}
In the case of an ideal measurement and perfectly indistinguishable photons, the visibility is connected to the PNC via
\begin{equation} \label{eq:def_visibility}
    v = \frac{|{\rho}_{0,1}|^2}{{\rho}_{1,1}} \,.
\end{equation}
In the MZI used in the experiment, the interference of subsequent single photons takes place. Phase scrambling between subsequent emission events leads to a further reduction of the visibility in addition to the aforementioned phonon and loss effects. It should be kept in mind that a vanishing visibility in the experiment can therefore result either from vanishing PNC, phase scrambling, or a combined effect of both. 

\subsection*{Theoretical expectations}
We perform theoretical simulations to estimate the PNC for both reX and stiX for a quantum dot modeled as a four-level system driven by a (classical) laser field  and coupled to discrete photon modes. The simulation considered the presence three photon modes, however, only the zero and one modes were found to be noticeably populated. We further account for coupling with acoustic phonons within a numerically exact path integral formalism \cite{Barth2016}. In addition, we include  relaxation between the quantum dot states accounting for photons not being emitted into the relevant modes (see Methods section for details of the model and calculation). We assume an ideal detection, i.e., no phase scrambling and perfect indistinguishability and  model the visibility via Eq.~\eqref{eq:def_visibility}.

\begin{figure*}[t]
    \centering
    \includegraphics[width=\linewidth]{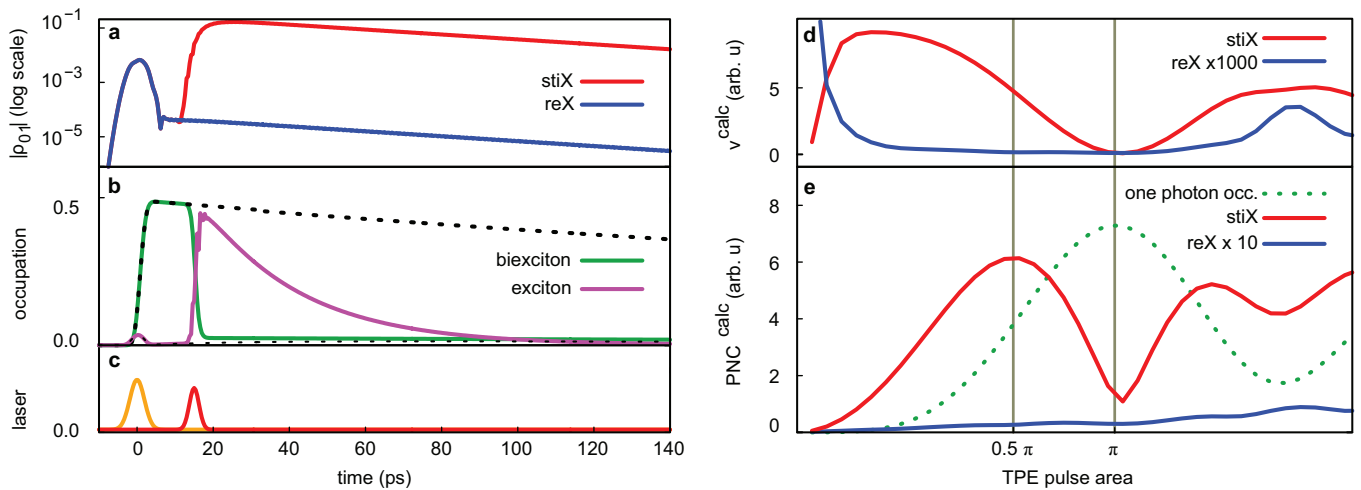}
    \caption{\textbf{Theoretical predictions}: Left: Dynamics of the four-level system coupled to two photon modes including phonons and losses calculated via a numerically exact path integral formalism. The exciting laser pulses with the TPE pulse (orange) and stimulating pulse (red) shown in (\textbf{c}). The occupation of the biexciton and exciton state for stiX are displayed in (\textbf{b}) with the dashed line indicating the behaviour for reX. The PNC for stiX (red) and reX (blue) is displayed in (\textbf{a}). Note the logarithmic scale. During the stimulating pulse the exciton becomes occupied resulting in a rise of the PNC. Right: Time-integrated coherence $\text{PNC}^{\text{calc.}}$ (\textbf{e}) and visiblity $v^{\text{calc}}$ (\textbf{d}) as a function of TPE pulse area for both reX (blue,magnified) and stiX (red). The TPE pulse areas of $\pi$ and $\pi/2$ are marked by vertical lines. The time-integrated occupation of the one-photon Fock state $\text{occ}^{\text{calc}}$ is shown as a green dashed line. Due to the relaxation process the PNC is almost lost in the reX case. In stiX, we find that the PNC is controlled via the TPE pulse area.} 
\label{fig:theory}
\end{figure*}

The time evolution of the biexciton and the $|x_H\rangle$ exciton occupation together with the PNC $|\rho_{0,1}|$ is shown in Fig.~\ref{fig:theory}\textbf{a,b}. Both schemes start with excitation from the ground into the biexciton state induced by a Gaussian-shaped laser pulse with a TPE pulse area of $\pi/2$. In the reX scheme, the biexciton state then relaxes into the exciton states via the emission of photons. However, these photons are at a different wavelength and therefore ignored. The exciton state is transiently occupied because it rapidly generates the desired photon relaxing further into the ground state. The corresponding PNC (blue curve in Fig.~\ref{fig:theory}\textbf{a}) is  almost vanishing, too, because of the incoherent biexciton-exciton relaxation destroying the electronic coherence. The remaining PNC can be traced back to deviations from the ideal case caused by phonon interaction, radiative losses, as well as relaxation into other (undesired) states in the quantum dot. 

In contrast, in the stiX scheme, the stimulating pulse brings the biexciton coherently into $|x_H\rangle$ by the application of a $\pi$-pulse resonant with the $|xx\rangle \to |x_H\rangle$ transition as evidenced in Fig.~\ref{fig:theory}\textbf{b}. The small oscillations on top of the population-exchange result from the off-resonant driving of the complementary transition $|x_H\rangle \to |g\rangle$. Because the transition to the exciton state $|x_H\rangle$ is coherent, the electronic coherence, which translates to the PNC, is preserved. Accordingly, in Fig.~\ref{fig:theory}\textbf{a} (red curve), we see that as soon as the stimulating pulse sets in, the PNC becomes very high. In other words, a timed stimulation preparation of the exciton state recovers the PNC that is lost in the reX scheme. 

By controlling the electronic coherence through the pulse areas of the exciting pulses, we can thus manipulate the PNC. To ensure the best comparability, we fix the stimulating pulse to a $\pi$ pulse and vary the pulse area of the TPE pulse, which results in an oscillating coherence as shown in Fig.~\ref{fig:intro}\textbf{c}. The time-integrated occupation of the one-photon state $\text{occ}^{\text{calc}}$ follows the Rabi rotations of the biexciton. We have checked that under the present conditions, the higher Fock states always have negligible occupations. The highest coherence is expected for pulse areas $(2n+1)\pi/2$, where also the electronic coherence is maximal. However, due to the incoherent relaxation process from the biexciton into the exciton, in reX the PNC is close to zero for all pulse areas as expected. This is confirmed by the numerical results in Fig.~\ref{fig:theory}\textbf{e}. Only for large pulse areas, detrimental processes due to phonons or losses lead to some residual PNC. Accordingly, for the reX scheme also the measured visibility $v^{\text{calc}}$ in Fig.~\ref{fig:theory}\textbf{d} is vanishing.

For stiX, the coherence is preserved and we find an oscillating behaviour as a function of the TPE pulse area with maxima of the PNC occurring for pulse areas with $(2n+1)\pi/2$ and minima for pulse areas $n\pi$. Ideally, PNC should be zero for pulse area $n\pi$. In the full simulation including finite pulse lengths and losses, the TPE pulse does not fully invert the system, leading to a residual PNC even for a TPE $\pi$-pulse. The visibility behaves differently: While a clear minimum at $\pi$ is recovered, the PNC is not maximal at $\pi/2$. Instead, due to its definition, $v^{\text{calc}}$ increases for the even smaller TPE pulse areas. Still, compared to reX, visibility for stiX shows a strong dependence on the TPE pulse area. 

\subsection*{Experimental data}

\begin{figure*}[ht]
    \includegraphics[width=1\linewidth]{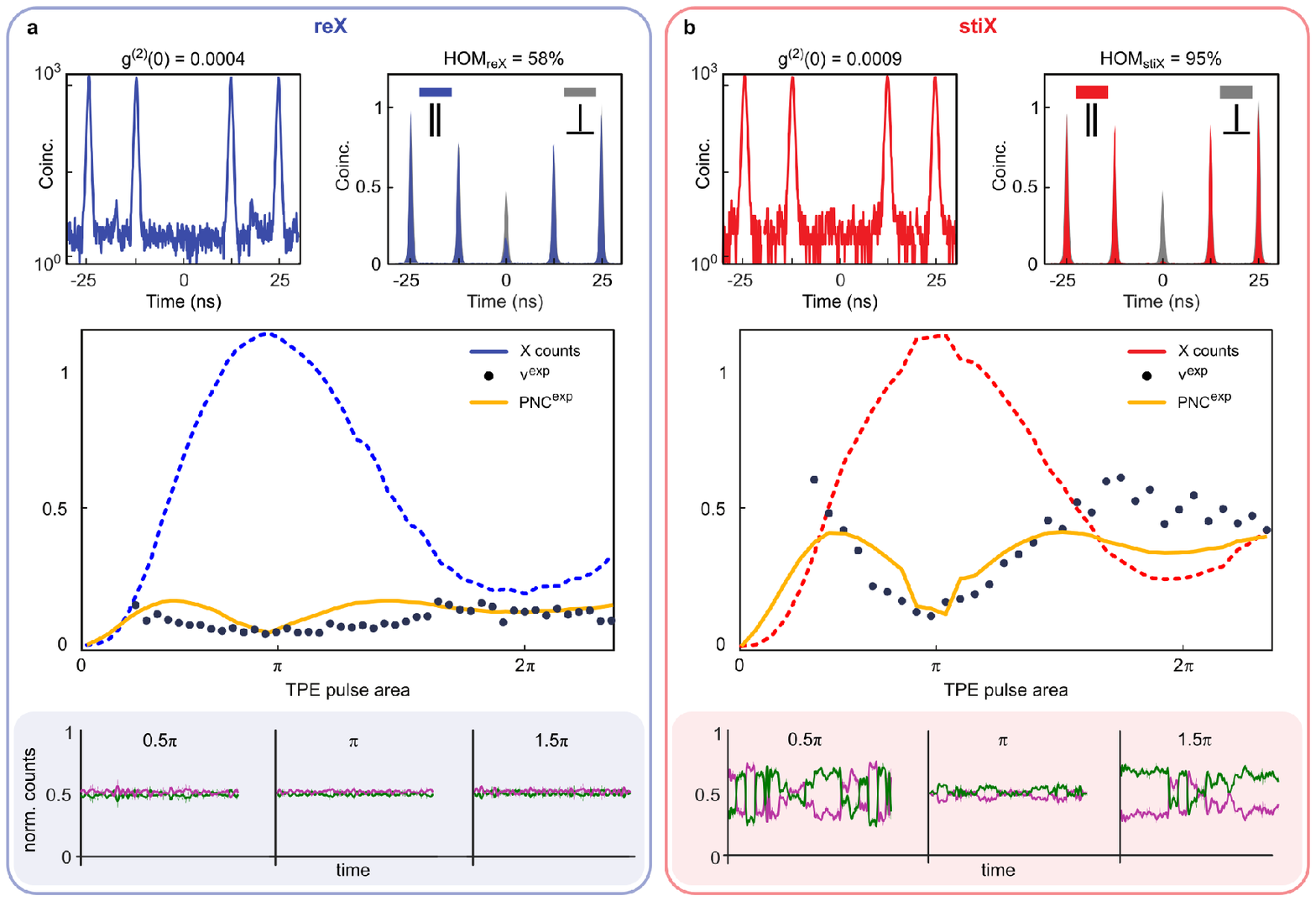}
    \caption{\textbf{Controlled generation of PNC on a single quantum dot for reX (\textbf{a}, left) and stiX (\textbf{b}, right)}: \textbf{Top panel}: Measured purity of the generated single photons. For $g^{(2)}_{\text{reX}}(0)$ (blue), the TPE pulse power is kept at $\pi$-power, and stim pulse is absent. For $g^{(2)}_{\text{stiX}}(0)$(red), TPE and stim pulses are kept at $\pi$-power. For $\text{HOM}_{\text{reX}}$, blue and gray shaded curves represent HOM coincidences recorded for parallel and orthogonal polarizations respectively, for $\pi$ TPE pulse area. For $\text{HOM}_{\text{stiX}}$, red and gray shaded curves represent HOM coincidences recorded for parallel and orthogonal polarizations respectively, TPE and stim pulses are kept at $\pi$-power. 
     \textbf{Middle panel}: Extracted visibilities $v^{\text{exp}}$ (dark-blue dots) and the reconstructed $\text{PNC}^{\text{exp}}$ (yellow) at different TPE pulse areas alongside the measured X counts (blue curve for $\text{X}_{\text{reX}}$ and red curve for $X_{\text{stiX}}$). The X counts are normalized to their respective values at $\pi$-power.  \textbf{Bottom panel}: Exemplary time traces recorded at the two detector outputs of the PNC setup at TPE pulse areas of $0.5\pi$, $\pi$, and $1.5\pi$.}
    \label{fig:experiments}
\end{figure*}

We perform the reX and stiX experiments to test the theoretical prediction on a single quantum dot in our setup displayed in Fig.~\ref{fig:intro}\textbf{d}. We note that in stiX we fix the time delay between the TPE pulse and stimulating pulse to \SI{7}{\pico\second}, where the photon count is maximal. A detailed description on the experiment is provided in the Methods Section. 

We start by quantifying the photon properties for reX and stiX, at various powers (see SI Table S1) measuring the single photon purity in a Hanbury Brown and Twiss (HBT) setup and the indistinguishability via Hong-Ou-Mandel (HOM) measurements. At $\pi$ power of both reX and stiX we validate that the generated photons have high purity with $g^{(2)}_{\text{reX}}(0) = 0.0004(1)$ and $g^{(2)}_{\text{stiX}}(0) = 0.0009(1)$. 
For the indistinguishability, the HOM visibility reaches only 58(3)\% under reX, while for stiX it increases to 95(6)\%, in line with previous observations \cite{sbresny2022stimulated,yan2022double,wei2022tailoring}. These results already underline that stiX is a advantageous scheme compared to reX. 

We then sweep the TPE pulse area under reX and stiX yielding Rabi rotations for the exciton (X) photon counts (blue and red curves in Fig.~\ref{fig:experiments}) and investigate the PNC. For each TPE pulse power, we analyze the spectrally filtered X photons using a phase-evolving MZI \cite{loredo2019generation}. Its outputs are simultaneously recorded with two avalanche photodiodes (APDs) for \SI{20}{\second} each. In the bottom panel of Fig.~\ref{fig:experiments}, we display exemplary time traces (denoted by green and magenta curves, representing the two detector outputs of the MZI, see Fig.~\ref{fig:intro}\textbf{d}) at TPE powers $0.5\pi$, $1\pi$, $1.5\pi$. From the time traces, we compute the visibility according to Eq.~\eqref{eq:def_visibility} from the normalized detector counts taking the average of the two detectors as $v^{\text{exp}}=(v_1+v_2)/2$.

The visibility $v^{\text{exp}}$ as a function of pulse area is displayed alongside the respective Rabi rotations in Fig.~\ref{fig:experiments} as black dots. Under reX, the visibility $v^{\text{exp}}$ is vanishing for all TPE pulse areas and no clear dependence is found. This is in agreement with the exemplary time traces (displayed in the bottom panel of Fig. \ref{fig:experiments}\textbf{a})., where indeed no oscillations are seen for different pulse areas   

In contrast, for stiX, the PNC shows a more interesting behaviour: in the exemplary time traces of the MZI outputs (displayed in the bottom panel of Fig. \ref{fig:experiments}\textbf{b}), we observe clear oscillations for TPE powers $0.5\pi$, and $1.5\pi$ and almost no oscillations at $\pi$. Accordingly, the visibilities vary from $\approx \,$0.6 at TPE power $0.5\pi$ to being minimal at $\pi$ and then rise again until $1.5\pi$. 

From the visibilities, using the formalism from Ref.~\cite{loredo2019generation}, we extract the $\text{PNC}^{\text{exp.}}$ shown as the yellow line in Fig.~\ref{fig:experiments}. The data clearly confirms the trend expected from the theory: We find minima of PNC when exciting with TPE pulses of pulse area $n\pi$ and maxima at $(2n+1)\pi/2$. This behaviour is evident in stiX, while in reX only a small modulation is found. 

Hence, we conclude that the PNC is negligible in reX, while in stiX we have tuneable PNC controlled via the TPE pulse area.

\section{Discussion}

We now set our results in the context of finding the optimal photon source for high-security quantum networks. As indicated before, purity, indistinguishability, and PNC are the key parameters that must be known when choosing an excitation scheme. We have shown that reX generates high-purity photons, while indistinguishability and PNC are low, and also, if filtering only a single polarization, the photon output is reduced. Looking back at Fig.~\ref{fig:QKD}\textbf{a}, we find that reX produces photons in the bottom left corner with low PNC and low indistinguishability, which limits the amounts of applicable protocols.

Within stiX, photons with high purity and high indistinguishability are generated. More importantly, the PNC can be controlled via the pulse area. If the TPE power is set to $(2n+1)\pi/2$, one obtains high PNC, enabling protocols in the top right corner of the diagram in Fig.~\ref{fig:QKD}\textbf{a}. By changing the pulse area to $n\pi$, the PNC is minimal which allows performing protocols in the top left corner of the diagram in Fig.~\ref{fig:QKD}\textbf{a}. For all TPE powers, stiX is suitable for protocols that require a high indistinguishability. Besides power control, the time delay of the stimulating pulse also controls the PNC (see SI Section \ref{sec:SI_timecontrol}). The largest PNC is obtained when the time separation between the TPE and the stimulating pulse is optimal.


In summary, we showed a controlled generation of single photons with variable degrees of PNC as well as high purity, high indistinguishability, and high brightness via a stimulated two-photon excitation. This is a big step forward towards the realization of secure quantum networks based on single photons.

\section{Methods}
\subsection{Theoretical model}
For the theoretical modelling, we set up the Hamiltonian consisting of the quantum dot system $\hat{H}^{\text{QD}}$, the out-coupling to two-photon modes $\hat{H}^{\text{photon}}$, the excitation of the TPE $\hat{H}^{\text{TPE}}$ and the stimulating laser pulse $\hat{H}^{\text{stim}}$, as well as the coupling to phonons 
\begin{equation}
    \hat{H} = \hat{H}^{\text{QD}} + \hat{H}^{\text{photon}} +  \hat{H}^{\text{TPE}} +\hat{H}^{\text{stim}}+ \hat{H}^{\text{phonon}}\,.
    \label{equ:full_hamiltonian}
\end{equation}
In addition, we consider radiative decay and losses by a Lindblad operator $\cal{L}$. In the following, we describe the individual terms in detail.

The quantum dot is modeled using four states (see also Fig.~\ref{fig:intro}\textbf{b}) denoted by $\vert g\rangle$ as the ground state, $\vert x_H\rangle$ and $\vert x_V\rangle$ as the two excitons and $\vert xx\rangle$ as the biexciton. The ground-state energy is set to zero, while both excitons have the same energy $\hbar\omega_x$, i.e., no fine-structure splitting is assumed. The biexciton has a binding energy $E_B$ such that its energy is given by $\hbar\omega_{xx} = 2\hbar\omega_x - E_B$.
\begin{equation}
    \begin{split}
        \hat{H}^{\text{QD}} = 
        & \hbar\omega_x \left(\vert x_H\rangle\langle x_H\vert + \vert x_V\rangle\langle x_V\vert \right)+  \hbar \omega_{xx}\vert xx\rangle\langle xx\vert 
    \end{split}
\end{equation}

The quantum dot is coupled to two photon modes with polarisations $V$ and $H$ for the out-coupling of the photons, similar to positioning the quantum dot in a photonic cavity. We model the photon modes by the Fock states $\vert n_H\rangle$ and $\vert n_V\rangle$ with the frequency $\omega_c$ via the  annihilation (creation) operators $\hat{a}_{H/V}$($\hat{a}^\dagger_{H/V}$). The photonic modes are coupled to the quantum dot transitions with the same strength via the coupling constant $\hbar g = 0.05\text{ meV}$, yielding
\begin{equation}
    \begin{split}
        \hat{H}^{\text{photon}} =& \hbar\omega_c\left(\hat{a}^{\dagger}_H\hat{a}_H + \hat{a}^{\dagger}_V\hat{a}_V \right)\\
        +& \hbar g \, \hat{a}_H \left( 
         \vert x_H\rangle\langle g\vert +
         \vert xx\rangle\langle x_H\vert\right) + h.c.\\
        +& \hbar g \, \hat{a}_V \left(
        \vert x_V\rangle\langle g\vert 
        +\vert xx\rangle\langle x_V\vert \right) + h.c. \\
        =& \hat{H}^{\text{photon}}_0 + \hat{H}^{\text{photon}}_{\text{coupl.}}.
    \end{split}
\end{equation}
We use the Hamiltonian in a rotating frame with ${\omega=\omega_l=\omega_x - E_B/(2\hbar)}$, which corresponds to the frequency of the TPE laser pulse. With this, the QD-photon Hamiltonian has the form
\begin{equation}
    \begin{split}
        \hat{H}^{\text{QD-photon}} =& 
         \hbar\Delta\omega_{x-l}\left(\vert x_H\rangle\langle x_H\vert + \vert x_V\rangle\langle x_V\vert \right) \\
        +& (\hbar 2\Delta\omega_{x-l} - E_B) \vert xx\rangle\langle xx\vert \\
        +&\hbar\Delta\omega_{c-l}\left(\hat{a}^\dagger_H\hat{a}_H + \hat{a}^\dagger_V\hat{a}_V\right) \\
        +& \hat{H}^{\text{photon}}_{\text{coupl.}}.
    \end{split}
\end{equation}
The index convention of frequency differences is chosen such that the second index is subtracted from the first, e.g., $\Delta\omega_{x-l} = \omega_x - \omega_l$. 
We choose the photon mode to be resonant with the quantum dot transition from the ground to the excited state, i.e., $\hbar\Delta\omega_{c-x}=\mbox{0\text{ meV}}$. 

The TPE is modelled by an external classical laser field with diagonal polarization in dipole and rotating wave approximation. We consider a resonant TPE process and accordingly set the detuning $\Delta\omega_{x-l} = E_B/(2 \hbar)$. With this, the Hamiltonian reads
\begin{equation}
    \begin{split}
        \hat{H}^{\text{TPE}}(t) = -\frac{\hbar}{2} f^{\text{TPE}}(t) &( \vert g\rangle\langle x_H\vert + \vert g\rangle\langle x_V\vert \\
        &+ \vert x_H\rangle\langle xx\vert + \vert x_V\rangle\langle xx\vert + h.c. ).
        \end{split}
\end{equation}
Here, $f^{\text{TPE}}(t)$ denotes the instantaneous Rabi frequency as given by the product of dipole moment and electric field. We use Gaussian pulses
\begin{equation}
    f^{\text{TPE}}(t) = \frac{ \Theta_{\text{TPE}}}{\sqrt{2\pi}\,\sigma_{\text{TPE}}}\text{e}^{-\frac{t^2}{2\sigma_{\text{TPE}}^2}},
\end{equation}
with the pulse area $\Theta_{\text{TPE}}$ and the pulse width $\sigma_{\text{TPE}}$. We assign the TPE pulse area $\pi$ in the plot (cf. Fig.~\ref{fig:theory} to the one which results in the first maximum of the biexciton occupation and the TPE pulse area of $\pi/2$ to the first maximum of the electronic coherence. In the calculations, these values were determined numerically.

We describe the stimulating laser with the same approximations, but assume it to be horizontally polarized. Its frequency is set to match the $\vert xx \rangle \to \vert x_H\rangle $ transition, such that 
\begin{equation}
    \hat{H}^{\text{stim}}(t) = -\frac{\hbar}{2}f^{\text{stim}}(t)\,
    \left[\text{e}^{i\Delta\omega_l^\text{stim}t}
    \left(\vert g\rangle\langle x_H\vert + \vert x_H\rangle\langle xx\vert \right)\right] + h.c. .
\end{equation}
Here, $\Delta\omega_l^\text{stim} = \omega_l^\text{stim}-\omega_l=-E_B/(2\hbar)$. The stimulating laser's envelope function $f^{\text{stim}}$ is delayed by a time $\Delta t$ compared to the TPE laser. We also assume a Gaussian envelope for the stimulating pulse 
\begin{equation}
    f^{\text{stim}}(t) = \frac{\Theta_\text{stim}}{\sqrt{2\pi}\,\sigma_\text{stim}}\text{e}^{-\frac{(t - \Delta t)^2}{2\sigma_\text{stim}^2}}.
\end{equation}
with the pulse area $\Theta_\text{stim}$ and the pulse length $\sigma_\text{stim}$. Here, a "$\pi$-pulse" refers to a full inversion of the resonantly driven transition for ideal conditions (without losses/phonons).

In addition we consider the coupling to longitudinal-acoustic (LA) phonons via the deformation potential coupling. Here, $\hat{b}_\mathbf{k}$ ($\hat{b}^\dagger_\mathbf{k}$) annihilates (creates) a phonon of mode $\mathbf{k}$ with energy $\omega_{\mathbf{k}}$. We consider the typical pure-dephasing type coupling in the standard Hamiltonian \cite{Besombes2001,Machnikowski2004}
\begin{equation}
    \hat{H}^{\text{phonon}} = \hbar\sum_{\mathbf{k}}\omega_{\mathbf{k}}\hat{b}^\dagger_{\mathbf{k}}\hat{b}_{\mathbf{k}} + \hbar \sum_{{\mathbf{k}},S}\left( \gamma_{\mathbf{k}}^S\hat{b}_{\mathbf{k}}^\dagger + \gamma_{\mathbf{k}}^{S^*}\hat{b}_{\mathbf{k}}\right)\vert S\rangle\langle S\vert,
\end{equation}
coupling each mode $\mathbf{k}$ to the quantum dot state $\vert S\rangle$, where $S\in \{x_H,x_V,xx\}$.  The coupling constant $\gamma_{\mathbf{k}}^S$ and the material parameters are taken to be the same as in Ref.~\cite{Barth2016}.

Both, cavity and quantum dot, are subject to losses into the free photonic field outside of the cavity. These losses are described by Lindblad-superoperators, affecting the density operator $\hat{\rho}$
\begin{equation}
    \mathcal{L}_{\hat{O}, \delta}[\hat{\rho}] = \delta\left( \hat{O}\hat{\rho}\,\hat{O}^\dagger - \frac{1}{2}\left[\hat{\rho}, \hat{O}^\dagger \hat{O}\right]_+\right),
\end{equation}
where $\hat{O}$ is an operator, $\delta$ a rate and $[.,.]_+$ the anti-commutator. We assume that the decay processes of the quantum dot take place with rate $\gamma$ and losses of the photonic modes go with the rate $\kappa$, such that Lindblad-superoperators are
\begin{equation}
    \begin{split}
        \mathcal{L}[\hat{\rho}] &:= \mathcal{L}_{\hat{a}_H,\kappa}[\hat{\rho}] + \mathcal{L}_{\hat{a}_V,\kappa}[\hat{\rho}] \\
        &+ \mathcal{L}_{\vert g\rangle\langle x_H\vert,\gamma}[\hat{\rho}] + \mathcal{L}_{\vert g\rangle\langle x_V\vert,\gamma}[\hat{\rho}] \\
        &+ \mathcal{L}_{\vert x_H\rangle\langle xx\vert,\gamma}[\hat{\rho}] + \mathcal{L}_{\vert x_V\rangle\langle xx\vert,\gamma}[\hat{\rho}] \,.
        \end{split}
\end{equation}
The rates are chosen such that we are in the weak coupling regime.

With the Hamiltonian and the Lindbladian terms we calculate the dynamics of the system states via the Liouville-von Neumann equation
\begin{equation} \label{eq:TheoryVonNeumannEq}
    \frac{\text{d}}{\text{d}t}\hat{\rho} = -\frac{i}{\hbar}\left[\hat{H}(t), \hat{\rho}\right] + \mathcal{L}[\hat{\rho}].
\end{equation}
As initial state we assume that the quantum dot is in its ground state and no photonic excitation exists. For the numerical integration, we use a numerically complete path-integral method, which is described in Refs. \onlinecite{Barth2016,Cygorek2017} and the parameters from Tab. \ref{tab:theoSystemParameters}, to solve Eq.~\eqref{eq:TheoryVonNeumannEq}.

\begin{table}
    \caption{Parameters used in the simulation. Material parameter are taken as in Ref. ~\cite{Barth2016}.}
    \label{tab:theoSystemParameters}
    \begin{tabular}{lc|c}
        \hline
        QD-cavity detuning &$\hbar\Delta\omega_{c-x}$ &  0 meV\\
        QD-laser detuning &$\hbar\Delta\omega_{x-l}$ & 2 meV\\
        detuning stim. pulse &$\hbar\Delta\omega_l^{\text{stim}}$ & 2 meV \\
        duration stim. pulse &FWHM$_{\text{stim}}$ & 3 ps\\
        duration TPE pulse& FWHM$_{\text{TPE}}$ & 4.5 ps\\
        delay between pulses &$\Delta t$ & 15 ps \\
        QD-cavity coupling &$\hbar g$ & 0.05 meV \\
        Binding energy &$E_B$ & 4 meV\\
        cavity loss rate &$\kappa$ &  0.577 ps$^{-1}$\\
        QD loss rate &$\gamma$ &  0.001 ps$^{-1}$\\
        QD size& $a$& 3 nm\\
        temperature & $T$ & 1.5 K
    \end{tabular}
\end{table}

We obtain results for the full density matrix, from which we can obtain the reduced density matrices for the quantum dots $\rho^{\text{QD}}_{S,S'} $ with $S\in \{g, x_H,x_V,xx\}$ and for the photons $\rho^{\text{photon}}_{n_i,n'_i}$ with $i\in\{H,V \}$, by tracing out the other degrees of freedom. We are interested in the coherence $\rho_{0,1}=\rho^{\text{photon}}_{0_H,1_H}$. The absolute value of $\rho_{0,1}=\rho^{\text{photon}}_{0_H,1_H}$ is referred to as PNC.

As a measure for the overall PNC at a given pulse area, we introduce the time-integrated absolute value of the instantaneous PNC
\begin{equation}
 \text{PNC}^\text{calc} \propto \tilde{\rho}_{0,1} 
    = \int  |\rho^{\text{photon}}_{0_H,1_H}| dt   \,.
\end{equation}
$\text{PNC}^\text{calc}$ is the calculated quantity that corresponds with experimental quantity $\text{PNC}^{\text{exp}}$ below in Eq.~\ref{eq:vis2pnc}.

Analogously, we define the time-integrated occupation of the one-photon number states as 
\begin{equation}
    \text{occ}^\text{calc} \propto  \tilde{\rho}_{1,1}=\int \rho^{\text{photon}}_{1_H,1_H} dt    \,.
\end{equation}
We assume that the photonic space can be reduced to a two-level system consisting of $|0_H\rangle$ and $|1_H\rangle$. This is reasonable because the higher-order Fock states are not occupied. We then follow Ref. \cite{loredo2019generation} to calculate the visibility $v$ as measure in a MZI for a mixed state as
\begin{equation}
    v^{\text{calc}} = \frac{\tilde{\rho}_{0,1}^2}{\tilde{\rho}_{1,1}} \,.
\end{equation}
We stress that this is an estimate of the visibility, which does not account for the imperfection of the beam splitter, higher photon states, phase scrambling, or reduced indistinguishability. Nonetheless, we expect the qualitative behaviour to agree with the experiment.

\subsection{Experimental setup}
Our setup (Fig.~\ref{fig:intro}\textbf{d}) consists of a Ti:Sapphire laser source (Tsunami 3950, SpectraPhysics) producing \SI{2.7}{ps} pulses (measured as intensity autocorrelation FWHM), that is tuned to \SI{793}{nm}, enabling spectral shaping of both the TPE and stimulating (stim). pulses via two independent 4$f$ pulse shapers. 
The intensities of the TPE and stim pulses are individually controlled via electronic variable optical attenuators (VOA, V800PA, Thorlabs) and the arrival time of the stimulating pulse is precisely controlled via a fiber optic delay line (ODL-300, OZ Optics). 
The two beams are combined at a 10:90 beamsplitter near the optical window of a closed-cycle cryostat (base temperature \SI{1.5}{\kelvin}, ICEOxford) where the quantum dot sample is mounted on a three-axis piezoelectric stage (ANPx101/ANPz102, attocube systems AG). The two beams are focused on a single quantum dot with a cold objective (numerical aperture 0.81, attocube systems AG). 

Our sample consists of GaAs/AlGaAs quantum dots with exciton emission centered around 790$\,$nm grown by the Al-droplet etching method \cite{huber2017highly,da2021gaas}. The dots are embedded in the center of a lambda-cavity placed between a bottom (top) distributed Bragg reflector consisting of 9 (2) pairs of $\lambda/4$ thick Al$_{0.95}$Ga$_{0.05}$As/Al$_{0.2}$Ga$_{0.8}$As layers.  

The quantum dot emission is collected via the same path as the excitation, where the exciton (X) photons are spectrally separated from the scattered laser light  and phonon side-bands using a home-built monochromator equipped with two narrow-band notch filters (BNF-805-OD3, FWHM \SI{0.3}{\nano\meter}, Optigrate).
To improve the suppression of the reflected TPE pulse we employ a cross-polarized configuration in which two orthogonal linear polarizers on excitation and collection paths block any residual laser scattering. In fact, this would not be necessary for a sufficiently narrow laser spectrum, as the TPE energy is detuned from the exciton energy. 
To measure the spectra, collected photons are routed to a single-photon sensitive spectrometer (Acton SP-2750, Roper Scientific) equipped with a liquid Nitrogen cooled charge-coupled device camera (Spec10 CCD, Princeton Instruments). For lifetime measurements, we use an avalanche photodiode (SPAD, Micro Photon Device) together with time-tagging electronics. 

\textbf{Phase scan HOM setup}: 
To measure the indistinguishability, the filtered X photons are sent through a Mach-Zehnder Interferometer (MZI) with a path-length difference of \SI{12.5}{\nano\second}, to interfere with successively emitted photons from the quantum dot in a 50:50 fiber beam splitter (TW805R5A2, Thorlabs) for HOM measurement. 
The two output ports of the fiber beam splitter are monitored by avalanche photodiodes (SPCM-NIR, Excelitas). The arrival times of the photons are recorded using a time tagger (Time Tagger Ultra, Swabian Instruments), and coincidence counting is employed to determine the correlation between the photons. In the HOM measurement, the polarization in both MZI arms is controlled individually, enabling a comparison between the co-polarized scenario with maximum indistinguishability and the cross-polarized situation with distinguishable photons to obtain the HOM visibility.

For PNC measurements, a phase shifter is placed into one of the arms of the unbalanced MZI. The phase shifter consists of a motorized rotation stage (ELL14K, Thorlabs) holding a half-wave plate positioned between two quarter-wave plates that are oriented orthogonally with respect to each other's fast axis. This arrangement effectively acts as a variable phase shifter for linearly polarised input light since:

\begin{equation}
    \begin{split}
        \mathrm{J(\theta)} &= \mathrm{QWP\left(\frac{\pi}{4}\right)} \cdot \mathrm{HWP\left(\theta\right)} \cdot \mathrm{QWP\left(-\frac{\pi}{4}\right)} \\ 
        &= -\frac{i}{2} \begin{bmatrix} 1 & -i \\ -i & 1 \end{bmatrix} \begin{bmatrix} \cos^2\theta - \sin^2\theta & 2\sin\theta\cos\theta \\ 2\sin\theta\cos\theta  & \sin^2\theta - \cos^2\theta  \end{bmatrix} \begin{bmatrix} 1 & i \\ i & 1 \end{bmatrix} \\ 
        & = \begin{bmatrix} 0 & e^{-i2\theta} \\  -e^{i2\theta} & 0\end{bmatrix}.
    \end{split}
    \label{equ:waveplates}
\end{equation}

Here $\theta$ is the orientation of the fast axis of the half-wave plate. 
By rotating the half-wave plate at a fixed speed, the phase in one of the arms is varied continuously without changing the polarization, while the phase in the other arm remains constant on the timescale of the rotation. The two arms are then recombined at the fiber beam splitter, where the interference occurs and photons are directed towards two separate single-photon detectors. The matching of the timing and relative polarization of the two arms was ensured by interfering the excitation laser with itself and maximizing the contrast, which yielded a visibility of 98$\,\%$.

\subsection{Extraction of the PNC from data} \label{sec:PNC_from_data}
We follow Ref.~\cite{loredo2019generation} to compute the PNC from the visibility. We remind that we only consider the $H$-polarized photons and stay in the approximation of the two-level system composed of the $|0\rangle$ and $|1\rangle$ Fock state. From the detector counts, we obtain the visibility $v^{\text{exp}}$, which is proportional to the occupation $\rho_{0,0}$. In the next step, we decompose the density matrix $\rho=\lambda \rho_\text{pure} + (1-\lambda) \rho_\text{mixed}$ into a part corresponding to a pure state and a part being a statistical mixture with the off-diagonal elements being zero. Note that we are only interested in the absolute value of the coherence and not in its phase. Following Ref.\cite{loredo2019generation}, the visibility can be approximated by $v\approx \lambda^2 \rho_{0,0} \sqrt{V_\text{HOM}}$ with $0\le \lambda \le 1$ and $V_\text{HOM}$ being the photon indistinguishability. Considering the slope of the visibility as a function of $\rho_{0,0} = (1-\rho_{1,1})$ allows us to extract $\lambda$ (see SI Section \ref{sec:SI_lambda}). Together with the knowledge of $\rho_{1,1}$ via the photon counts, we can estimate the PNC as
\begin{equation} 
    \text{PNC}^{\text{exp.}}  =\lambda \sqrt{\rho_{1,1}(1-\rho_{1,1})}\,.
    \label{eq:vis2pnc}
\end{equation}

\section{Acknowledgements}
The authors gratefully acknowledge insightful discussions with Stefan Frick, Robert Keil, Tommaso Faleo, Mathieu Bozzio and Serkan Ates. Nils Kewitz and Bhavana Panchumarthi supported the early phases of the experiment. YK, FK, RS, VR and GW acknowledge financial support through the Austrian Science Fund FWF projects W1259 (DK-ALM Atoms, Light, and Molecules), FG 5, TAI-556N (DarkEneT) and I4380 (AEQuDot). DAV and TH acknowledge financial support by the German Federal Ministry of Education and Research (BMBF) via projects 13N14876 (‘QuSecure’) and  16KISQ087K (tubLAN Q.0). TKB and DER acknowledge financial support from the German Research Foundation DFG through project 428026575 (AEQuDot). A.R. and SFCdS acknowledge the FWF projects FG 5, P 30459, I 4320, the Linz Institute of Technology (LIT) and the European Union's Horizon 2020 research, and innovation program under Grant Agreement Nos. 899814 (Qurope), 871130 (ASCENT+) and the QauntERA II Programme (project QD-E-QKD).
LMH, PW and JCL acknowledge financial support  from the European Union’s Horizon 2020 and Horizon Europe research and innovation programme under grant agreement No 899368 (EPIQUS), the Marie Skłodowska-Curie grant agreement No 956071 (AppQInfo), and the QuantERA II Programme under Grant Agreement No 101017733 (PhoMemtor); FWF through F7113 (BeyondC), and FG5 (Research Group 5); from the Austrian Federal Ministry for Digital and Economic Affairs, the National Foundation for Research, Technology and Development and the Christian Doppler Research Association. 
For the purpose of open access, the author has applied a CC BY public copyright licence to any Author Accepted Manuscript version arising from this submission.


\section{Author contributions}
The experimental setup was built by Y.K., F.K., R.S., V.R., J.C.L., D.A.V., L.M.H., and the measurements were performed by Y.K., F.K., D.A.V. The numerical calculations were done by P.C.A.H. The sample was provided by C.S., S.F.CdS, A.R. The first draft of the manuscript was written by D.A.V., F.K., Y.K., P.C.A.H., V.R., D.E.R.. Conceptual work and supervision was done by G.W., P.W., V.M.A, D.E.R., V.R., J.C.L., T.H., A.R..
All authors discussed the results and were involved in writing the manuscript.

\bibliography{reference_noDOI.bib}



\clearpage

\onecolumngrid
\appendix
\begin{center}
    \Large{\textbf{Supplementary information}}
\end{center}

\section{Quantum Dot Sample}
The sample used contains GaAs/AlGaAs quantum dots (QDs) obtained by the Al-droplet etching method \cite{da2021gaas} and was grown by molecular beam epitaxy. The QD are embedded in the center of a $\lambda$-cavity placed between a bottom(top) distributed Bragg reflector consisting of 9(2) pairs of $\lambda/4$-thick $Al_{0.95}Ga_{0.05}As/ Al_{0.20}Ga_{0.80}As $layers with respective thickness of 69/60 nm. The QDs are placed between to $ \lambda/2$-thick $Al_{0.33}Ga_{0.67}As$ layers. The QD growth process starts by depositing 0.5 equivalent monolayers of Al in the absence of arsenic flux, which results in the self-assembled formation of droplets. During exposure to a reduced As flux, such droplets locally etch the underlying $Al_{0.33}Ga_{0.67}As$ layer, resulting in $\approx$\SI{9}{\nano\meter}-deep and $\approx$\SI{60}{\nano\meter} wide nanoholes on the surface. Then the nanoholes are filled with GaAs by depositing $\approx$\SI{1.1}{\nano\meter} of GaAs on the surface, followed by an annealing step of \SI{45}{\second}. The temperature used for the etching of the nanoholes was \SI{600}{\celsius}. The droplet self-assembly process results in QDs with random position and a surface density of about $2\times 10^{7} \si{\per\centi\meter\squared}$, suitable for single QD spectroscopy.

\section{stiX characterization and optimization}

To confirm that the collected photons indeed originate from stiX, several characterisation experiments are done, as presented in the following.

\begin{figure*}[!hbt]
    \centering
    \includegraphics[width=1\linewidth]{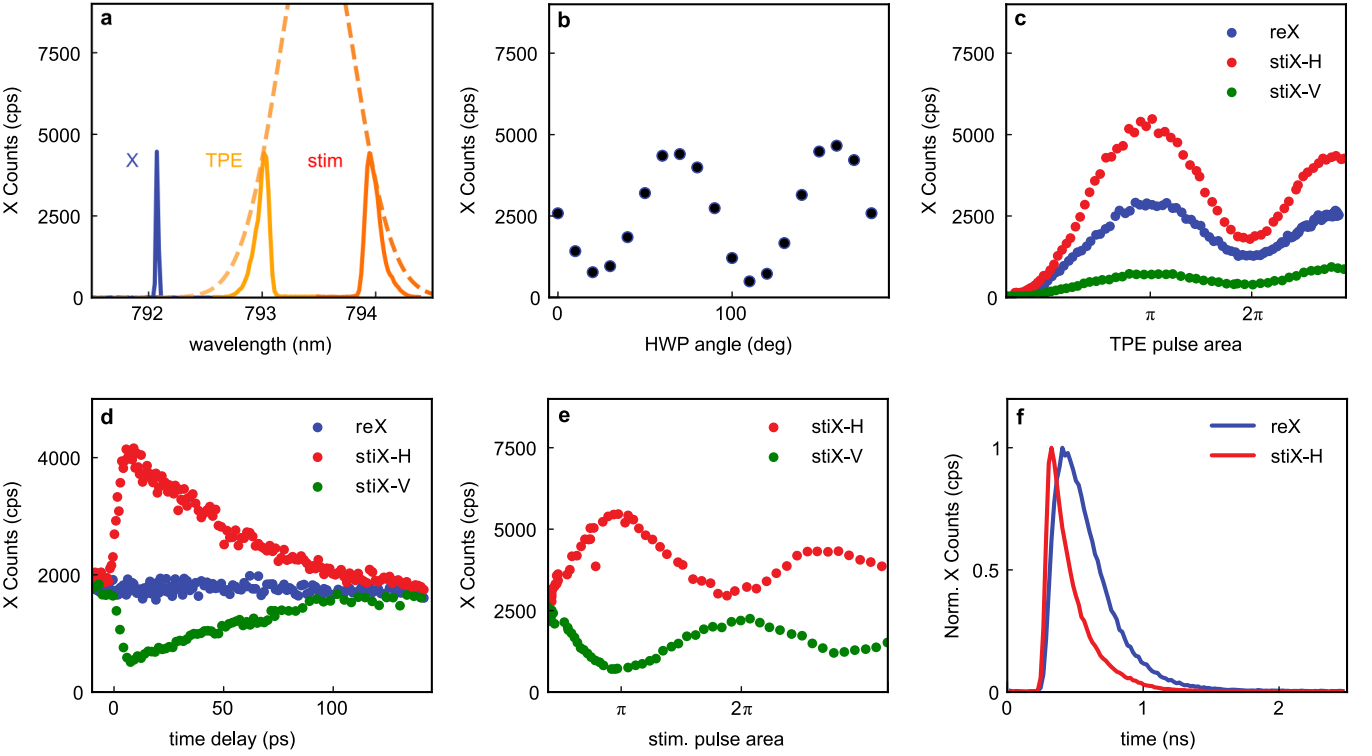}
    \caption{\textbf{stiX characterization}: (a) Pulse shaping and detuning during the experiment. TPE (light orange) and stim. (dark orange) pulses are spectrally cut from an initial picosecond pulse (dashed orange line).  (b) Integrated $X_H$ counts as a function of the stimulation pulse polarization tuned via an HWP (at optimal delay). (c) Measured two-fold $X_H$ photon enhancement (red dots) under optimal stiX conditions of time delay, power and polarization. The green dots represent the case where we suppress the collected polarisation. Blue dots represent the reX case. (d) Integrated $X_H$ photon counts as a function of the time separation of stim. pulse, following the TPE pulse (red dots). Blue dots represent the $X_H$ photon counts under reX. Green dots represent the drop in photon counts for a vertically polarized stim. pulse. (e) Integrated $X_H$ counts for varying stim. pulse power while keeping the TPE pulse at $\pi$ power. Red dots represent the H cascade, which is enhanced, while green dots represent the V-polarized cascade, which is suppressed. (f) Emission decay dynamics under stiX (red) compared to reX (blue) in a lifetime measurement. } 
    \label{fig:stiX}
\end{figure*}

Initially, we vary the TPE power, observing Rabi rotations to determine the TPE $\pi$ power. While maintaining the TPE pulses at this power, we introduce the stim. pulse tuned to the XX energy and scan their relative time delay (see Methods in the main manuscript) from negative (i.e., stim. pulse arrives before the TPE pulse) to positive while recording the generated X photons.
Optimum stiX is achieved when the stim. pulse arrives just after the TPE pulse has maximized the XX state occupation. Arriving earlier will not stimulate the de-excitation and a stim. pulse that arrives too late will allow the XX to already decay spontaneously. This explains the observed dependence of the X emission on the temporal delay between TPE and stim. pulses, a sharp increase for short delays and an exponential reduction of the stimulation effect that follows the XX lifetime. In Ref. \onlinecite{sbresny2022stimulated} the optimum delay is about $0.03 \times \tau_{XX} \approx 5\, \text{ps}$ which matches our observations (Fig.~\ref{fig:stiX}\textbf{d}). The red (stim. pulse has H polarization), green (stim pulse has V polarization) and blue (no stim. pulse) dots represent the different polarization cases respectively. At time delay $\approx$\SI{7}{\pico\second} we observe nearly two times the photon counts compared to the reX case for an H-polarized stim. pulse, confirming the successful stimulation. Note that, when we stimulate the V-polarization cascade, we are effectively suppressing the collected decay pathway, which shows as a drop in photon counts (see Fig.S\ref{fig:stiX}\textbf{d}, green dots).

As the stim. pulse removes the timing jitter from the exciton emission caused by the biexciton decay, one can also observe the effect in the X photon arrival time distribution (Fig.~\ref{fig:stiX}\textbf{f}). The standard reX exciton emission (blue line) exhibits an exponential rise time that corresponds to the XX decay time characteristic of the cascaded emission, before decaying exponentially. When adding the stim. pulse (red line) the timing jitter due to the random biexciton decay is removed and the rise time of the exciton vanishes. The exciton now decays right after the arrival of the stim. pulse. 


Once the time delay of the stim. pulse is optimized, one can turn to the polarization control. To achieve maximum enhancement, the polarization of the stimulation pulse that controls the emitted X polarization must match the polarization that is collected, while the orthogonal polarization leads to suppression. This is confirmed by rotating the stim. pulse polarization at optimum delay and power by the means of an HWP (figure \ref{fig:stiX}\textbf{b}). In reX, the polarization of the excitation pulse does not have an effect on the X emission polarization, as any linear polarization would generate the H and V cascades to the ground state. However, with the stim. pulse polarization, one can direct the emission cascade to arbitrary ratios of $X_{H}$ or $X_{V}$. We observe an oscillatory trend with respect to the HWP orientation, signifying that the $X_{H}$ intensity can be controlled via stiX polarization.
Note that, while the enhancement increases the emission by a factor of 2, the suppression does not reduce the counts completely. This is also observed in other works on stiX \cite{wei2022tailoring} and can be attributed to an imperfect preparation fidelity of the reX, non-ideal stim. pulse polarization control as well as anisotropies in the quantum dot that cause the polarization axes to not be fully orthogonal, such that the stim. pulse also couples with the other polarization axis \cite{krizhanovskii2005individual,belhadj2010impact}. 

Next, we investigate the power dependence of the stim. pulse by setting the TPE pulse to $\pi$-power and sweeping the stim. pulse power (figure \ref{fig:stiX}\textbf{e}). This is again done for the stim. pulse with horizontal (enhancement, red dots) and vertical (suppression, green dots) polarization. The coherent excitation and de-excitation of the quantum dot states is evidenced by the presence of Rabi rotations as a function of stim. pulse power consistent with previous works \cite{sbresny2022stimulated,wei2022tailoring,yan2022double}. While adding the stim. pulse depopulates the biexciton to enhance or suppress the exciton emission in one polarization, higher TPE powers can re-populate the biexciton state which compensates the effect of the stim. pulse. Thus, the emission also oscillates as a function of stim. pulse pump area.

Finally, we can quantify the photon enhancement at this optimal stiX condition. At first, we switch off the stim. pulse, and record the X photon counts upon TPE pulse power sweep, to observe the reX Rabi oscillations (Fig.~\ref{fig:stiX}\textbf{c}, blue dots). At $\pi$-power we observe $\approx$2600 counts. We now introduce the stim. pulse, with its polarization set to H, and perform the TPE power sweep. The red dots in Fig.~\ref{fig:stiX}c represent the recorded photon counts which, at $\pi$-power reach $\approx$5200 counts, clearly demonstrating the expected two-fold enhancement under optimal stiX conditions. Furthermore, we observe that once we switch to orthogonal polarization of the stim. pulse, the recorded photon counts are minimal, which represent the residual photon emission via the V-cascade (Fig.~\ref{fig:stiX}\textbf{c}, green dots).

Finally, the hallmark and main motivation of the stiX scheme is the improved indistinguishability that was already shown in the main text. In summary, all these observations consistently confirm the successful realization of the stiX scheme with two pulses spectrally cut from the same ps-laser. 

\section{Photon quality}
In Table \ref{tab:g2_comparison} we summarize the single-photon purities and indistinguishabilities measured under various excitation conditions of reX and stiX methods. To compute the $g^{(2)}(0)$, we first fit the recorded photon coincidences for a time interval of \SI{8}{\nano\second} around zero delay with a Gaussian function, and also the four side peaks (corresponding to time delays $\pm\SI{12.5}{\nano\second}$ and $\pm\SI{25}{\nano\second}$) to then calculate the ratio of the extracted areas. 
For calculating HOM visibility, we again fit the recorded photon coincidences for a time interval of \SI{6}{\nano\second} at zero delay for parallel and orthogonal configurations and compute the ratio of the extracted areas. 

\begin{table}[!hbt]
    \centering
    \caption{Single-photon purity, measured as $g^{(2)}(0)$, and indistinguishability (HOM Visibility $\mathrm{V_{HOM}}$) under various excitation pulse powers of reX, stiX and s-shell excitation. Uncertainties are extracted from the fitting procedure.}
    \begin{tabular}{c|c|c|c|c}
        Protocol & TPE power  & stim. power  & $g^{(2)}(0)$ & $\mathrm{V_{HOM}}$ \\
        \hline
        reX & $\pi$ & - & 0.0004(1) & 0.58(3) \\ \hline
        stiX & 0.25$\pi$ & $\pi$ &- & 0.85(3) \\
        stiX & 0.5$\pi$ & $\pi$ &-& 0.73(5)\\
        stiX & $\pi$ & $\pi$ & 0.0009(1) & 0.95(6)\\
        stiX & 2$\pi$ & $\pi$ & 0.0003(1) &-\\  \hline
        s-shell & -& - & 0.04(1) & 0.88(3)
    \end{tabular}
    
    \label{tab:g2_comparison}
\end{table}

%

\section{Quantum Efficiency}
\begin{figure*}[!hbt]
    \centering
    \includegraphics[width=1\linewidth]{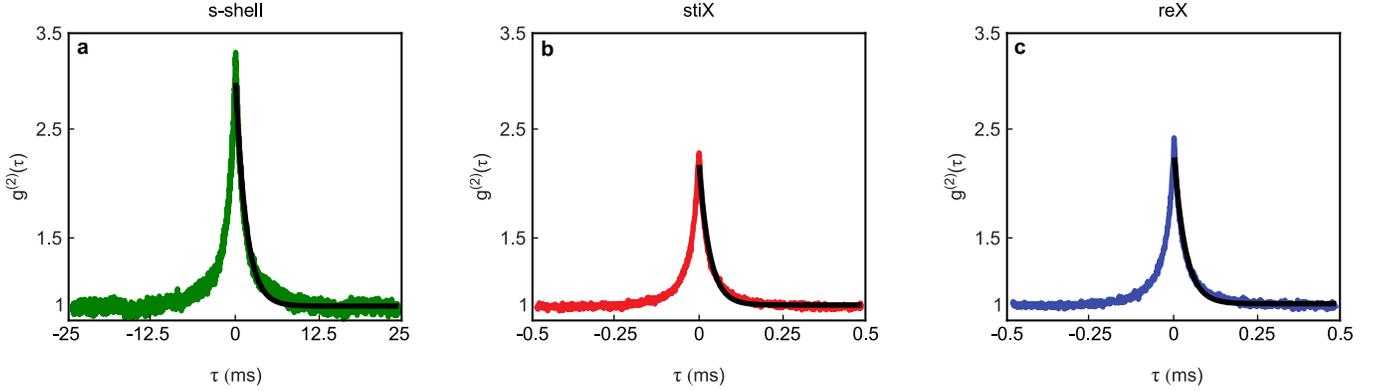}
    \caption{\textbf{Quantum efficiency}: Blinking behaviour of the quantum dot under different excitation schemes, as measured by long timescale $g^{(2)}(\tau)$. (a): resonant s-shell excitation, (b): stiX, and (c): reX . Black curve denote the fit.}
    \label{fig:blinking}
\end{figure*}

The blinking behavior of the  quantum dot varies depending on the type of excitation. In Fig.~\ref{fig:blinking}\textbf{(a)-(c)} we present the behaviour under resonant excitation, reX and stiX of the exciton state, on the same quantum dot, as measured under a long timescale $g^{(2)}(0)$, denoted here as $g^{(2)}_{\text{LT}}(0)$ \cite{munzberg2022fast,schimpf_quantum_2021}.  We fit a symmetric exponential function $g^{(2)}(\tau) = \text{A}*\exp(-|\tau/\tau_{\text{blinking}}|) + \text{B} $ (black solid lines) to extract a blinking timescale $\tau_{\text{blinking}}$ and $g^{(2)}_{\text{LT}}(0) = \text{A} + \text{B}$ . The extracted parameters are listed in Table \ref{tab:blinking}.

\begin{equation}
    \text{QE}=\frac{t_{\mathrm{on}}}{t_{\mathrm{off}}+t_{\mathrm{on}}}=\frac{1}{g^{(2)}_{\text{LT}}(0)}
\end{equation}

\begin{table}[!hbt]
    \centering
    \caption{Blinking behaviour analysis for different excitation techniques. Uncertainties are extracted from the fitting procedure.}
    \begin{tabular}{c|c|c | c}
        Protocol & $\tau_{\text{blinking}}$ / \si{\milli \second} &  $g^{(2)}_{\text{LT}}(0)$ & QE\\
        \hline
        s-shell &1.601(6)& 2.865(5) & 0.349(1)\\
        reX & 0.0350(3) & 2.247(7) & 0.445(1)\\
        stiX & 0.0326(3) & 2.184(7) & 0.458(1)
        
    \end{tabular}
    
    \label{tab:blinking}
\end{table}

We calculate that under resonant s-shell excitation, the quantum efficiency of the quantum dots is $\approx 0.35$. By introducing reX and stiX, we are able to achieve higher quantum efficiencies of $\approx 0.45$ and $\approx 0.46$, respectively.


\clearpage
\section{Time control of PNC} \label{sec:SI_timecontrol}
\begin{figure*}[!hbt]
   \centering
   \includegraphics[width=1\linewidth]{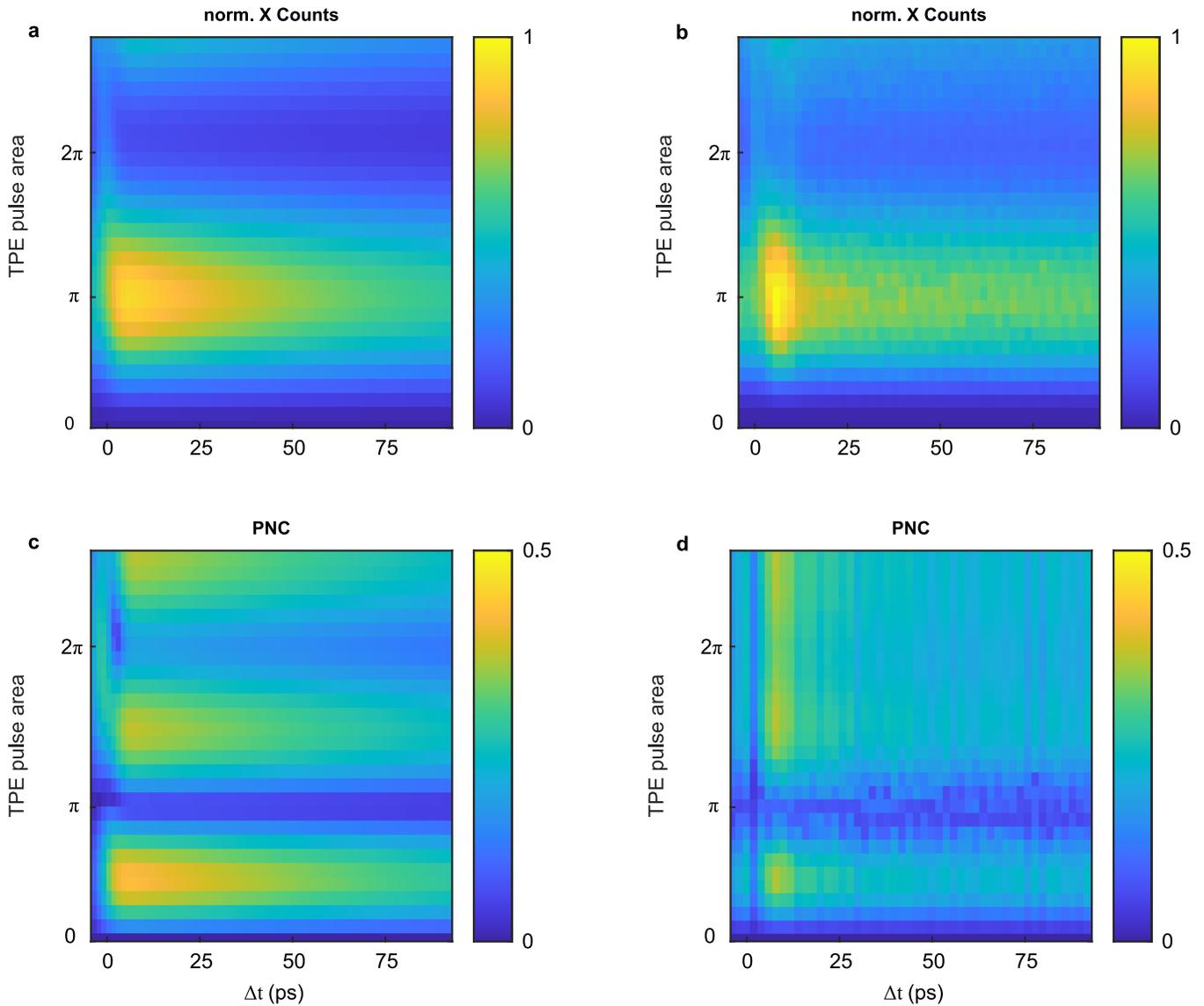}
   \caption{\textbf{stiX and PNC temporal behaviour}: (a) \textbf{Theory}: Mean photon number per excitation cycle depending on TPE pulse power and time seperation between TPE and stim pulses ($\Delta t$). Every column represents a TPE pulse power sweep and each row the corresponding stiX time delay scan, with stiX power set to $\pi$. (b)\textbf{Experiment}: Integrated X photon counts for the same parameters as in (a). (c) \textbf{Theory}: Simulated PNC of stiX. (d) \textbf{Experiment}: Qualitative (e.g. $V_{HOM}$ is constant for all $\Delta t$) extraction of PNC from the measurement.}
   \label{fig:stiX2Dmap}
\end{figure*}

Here, we investigate the control of PNC further using the stiX scheme parameters. Initially, we scan the time delay of the stim. pulse and record the $X_{\text{H}}$ counts for various TPE pulse powers. The resulting two-dimensional map of the recorded photon counts of $X_H$ photons as a function of TPE power and stim. pulse time delay is displayed in Figure~\ref{fig:stiX2Dmap}b. We observe that the photon counts reach the expected two-fold enhancement (compared to reX case, c.f. Figure \ref{fig:stiX}\textbf{c}) at a time delay $\approx$\SI{7}{\pico\second} when TPE pulse power reaches $\approx$1$\pi$. 
Subsequently, we perform the PNC experiment (see Methods) at every time delay and TPE power, and extract the visibilities (same as in Figure \ref{fig:experiments} in the main manuscript). We then compute the PNC (note that $V_\text{HOM}$ is assumed constant for all time delays) using the same procedure as explained in Section~\ref{sec:PNC_from_data} and \ref{sec:SI_lambda}.  
The resulting two-dimensional map of the computed PNC values for various time delays and TPE powers is displayed in Fig.~\ref{fig:stiX2Dmap}d.

To support our observations, we perform additional simulations on the mean expected photons per excitation cycle ($X_H$) and the  PNC. For computational simplicity, omit the two photon modes and the coupling to them, such that the Hamiltonian \eqref{equ:full_hamiltonian} reduces to 
$    \hat{H} = \hat{H}^{\text{QD}} +  \hat{H}^{\text{TPE}} +\hat{H}^{\text{stim}}+ \hat{H}^{\text{phonon}}$\,. 
This is motivated by the agreement between the PNC and the quantum dot coherence (see Fig.~\ref{fig:intro}b and Fig.~\ref{fig:theory}e) by time-integrating the respective elements and normalizing by the decay rate $\gamma_X$, i.e.:

\begin{equation}
    \begin{split}
         X_H & \approx \gamma_X \int |\rho_{X_H,X_H}|  dt \\
         \text{PNC} & \approx  \frac{\gamma_X}{2}\int |\rho_{g,X_H}| dt .
     \end{split}
\end{equation}

The results are displayed in Fig.~\ref{fig:stiX2Dmap}a and \ref{fig:stiX2Dmap}c, and show an excellent match with our experimental data.

\clearpage
\section{Estimation of $\lambda$ for PNC extraction from visibility} \label{sec:SI_lambda}

\begin{figure}[!hbt]
    \centering
    \includegraphics[width=0.5\textwidth]{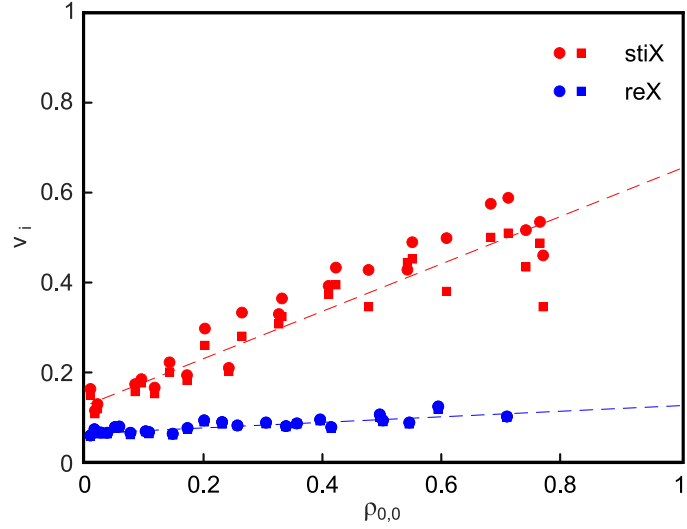}
    \caption{\textbf{Computing $\lambda$}: 
    Squares (circles) represent the measured visibilities $v_{1(2)}$ measured at APD1(2) respectively. Red and blue denote stiX and reX processes. Dashed lines are linear fits according to equation \ref{eq:lamb_from_vis}.} 
    \label{fig:lambda}
\end{figure}

We compute PNC based on Equation~\eqref{eq:vis2pnc}, which depends on $\lambda$, a quantity that signifies the purity of the state. Following the method in Ref. \onlinecite{loredo2019generation} we extract this parameter by fitting the measured visibility $v$ (c.f. Figure \ref{fig:experiments} middle panel), as 
\begin{equation}
    v\approx \lambda^2 \rho_{0,0} \sqrt{V_\text{HOM}} + v_0.
    \label{eq:lamb_from_vis}
\end{equation} 
Here $V_\text{HOM}$ is the measured single-photon indistinguishability, $v_0$ a residual visibility at $\pi$ power and $\rho_{0,0}$ is approximated by $\rho_{0,0} \approx 1- \text{N}$ with $\text{N}$ being the normalized photon counts measured while sweeping the TPE pulse power. This approximation is valid since the system does not include decay channels that do not end in the ground state and we treat the photon density matrix in a two-dimensional subspace. Fitting equation \eqref{eq:lamb_from_vis} we obtain $\lambda_{\text{stiX}} = 0.73(3)$ for stiX (displayed as red circles and squares, respectively indicating the measured visibilities from two detectors) and $\lambda_{\text{reX}} = 0.28(2)$ for reX. The results are displayed in Fig.~\ref{fig:lambda}. 

\clearpage

\end{document}